\documentclass{article}
\usepackage{cite}

\usepackage[caption=false,font=footnotesize]{subfig}

\usepackage{graphicx}
\usepackage{amssymb}
\usepackage{epstopdf}
\usepackage{url}

\usepackage{fixltx2e}
\usepackage{stfloats}

\pdfoutput=1

\newcommand{\Physarum}{ \emph{P.~polycephalum} }

\begin{document}

\title{Slime mould solves maze in one pass $\ldots$ assisted by gradient of chemo-attractants}

\author{Andrew Adamatzky
\thanks{University of the West of England, Bristol BS16 1QY, United Kingdom
Email: {andrew.adamatzky@uwe.ac.uk}}}

\markboth{Andrew Adamatzky}%
{Slime mould solves maze in one pass}

\maketitle

\begin{abstract}
Plasmodium of \emph{Physarum polycephalum} is a large cell, visible by unaided eye, which exhibits sophisticated patterns of foraging behaviour. The plasmodium's behaviour is well interpreted in terms of computation, where data are  spatially extended configurations of nutrients and obstacles, and results of computation are networks of protoplasmic tubes formed by the plasmodium. In laboratory experiments and numerical simulation we show that 
if plasmodium of \Physarum   is inoculated in a maze's peripheral channel and an oat flake (source of attractants) in a  
the maze's central chamber then the plasmodium  grows toward target oat flake and connects the flake with the site of original inoculation with a pronounced protoplasmic tube. The protoplasmic tube represents a path in the maze.  The plasmodium solves maze in one pass because it is assisted by a gradient of chemo-attractants propagating from the target oat flake. 

\vspace{0.5cm}

\end{abstract}

\section{Introduction}

 A typical strategy for a maze-solving is to explore  all possible passages, while marking visited parts, till the exit or a central chamber is found. This is what Shannon's electromagnetic mouse Theseus was doing in first ever laboratory experiment on solving maze by physical means~\cite{shannon_1952}. 
The task of maze search is time consuming for a single mobile computing device. Therefore with advent of unconventional computing paradigm scientists focused on uncovering physical, chemical and biological substrates which can solve maze in parallel. Thus in \cite{steinbock_1995} it is experimentally demonstrated that detecting superposition of excitation wave fronts propagating from source to destination, and from destination to source, in a maze filled with Belousov-Zhabotinsky (BZ) reaction mixture allows us to approximate a shortest path in the maze. Travelling excitation waves propagate to all channels of the maze thus examining the maze's structure in full. The  BZ maze solver does not represent path  by the medium's physical characteristics --- an external observer is required to reconstruct a shortest path from chemical waves' dynamics.  

In a gas-discharge maze-solver~\cite{reyes_2002} an electric field, generated between source and destination electrodes, explores a maze in parallel. The field is stronger along the shortest path between source- and destination-electrodes therefore channels of  the shortest path glow with high intensity. Such maze-solver has a volatile memory because the shortest path is visible as long as voltage to the electrodes is applied. A slime mould maze-solver reported in~\cite{nakagaki_2001a} is a fusion of BZ and gas-discharge approaches.
Plasmodium of \Physarum  is placed in several sites of maze at once. Initially the plasmodium develops a network of protoplasmic tubes spanning all channels of the maze, thus representing all possible solutions. Then oat flakes are placed in source and destination sites and the plasmodium enhances the tube connecting source and destination along the shortest path. A selection of shortest protoplasmic tube is implemented via interaction of propagating bio-chemical, electric potential and contractile waves in plasmodium's body. Not shortest or cul-de-sac branches are abandoned. 
Thus we can call prototype~\cite{nakagaki_2001a} as 'pruning plasmodium'. The largest protoplasmic tube (exactly its walls), which represents the shortest path, remains visible even when plasmodium ceases functioning. Therefore we can consider the slime mould maze-solver as having non-volatile memory (analogous to precipitating reaction-diffusion chemical processors). 

In laboratory prototypes of a  mobile droplet~\cite{lagzi_2010} and hot ice computer~\cite{adamatzky_hotice} a computation of shortest path in a maze is separated on two stages: computation of many-sources-one-destination set of paths and extraction of a shortest path from the  set of paths.  Many-sources-one-destination paths are computed by spreading acidity in~\cite{lagzi_2010} and propagating crystallisation pattern in~\cite{adamatzky_hotice}. A shortest path  is  selected and traced by a droplet travelling along pH gradient~\cite{lagzi_2010} or a virtual robot traversing crystallisation 
pattern~\cite{adamatzky_hotice}. In present paper we apply the two-stage computation in design of slime mould maze-solver which navigates a maze in one go, without exploring all possible options because the options are `explored' by  chemo-attractants emitted by destination oat flake.

\section{Experimental maze-solving}

\begin{figure*}[!t]
\centering
\subfloat[]{\includegraphics[width=0.49\textwidth]{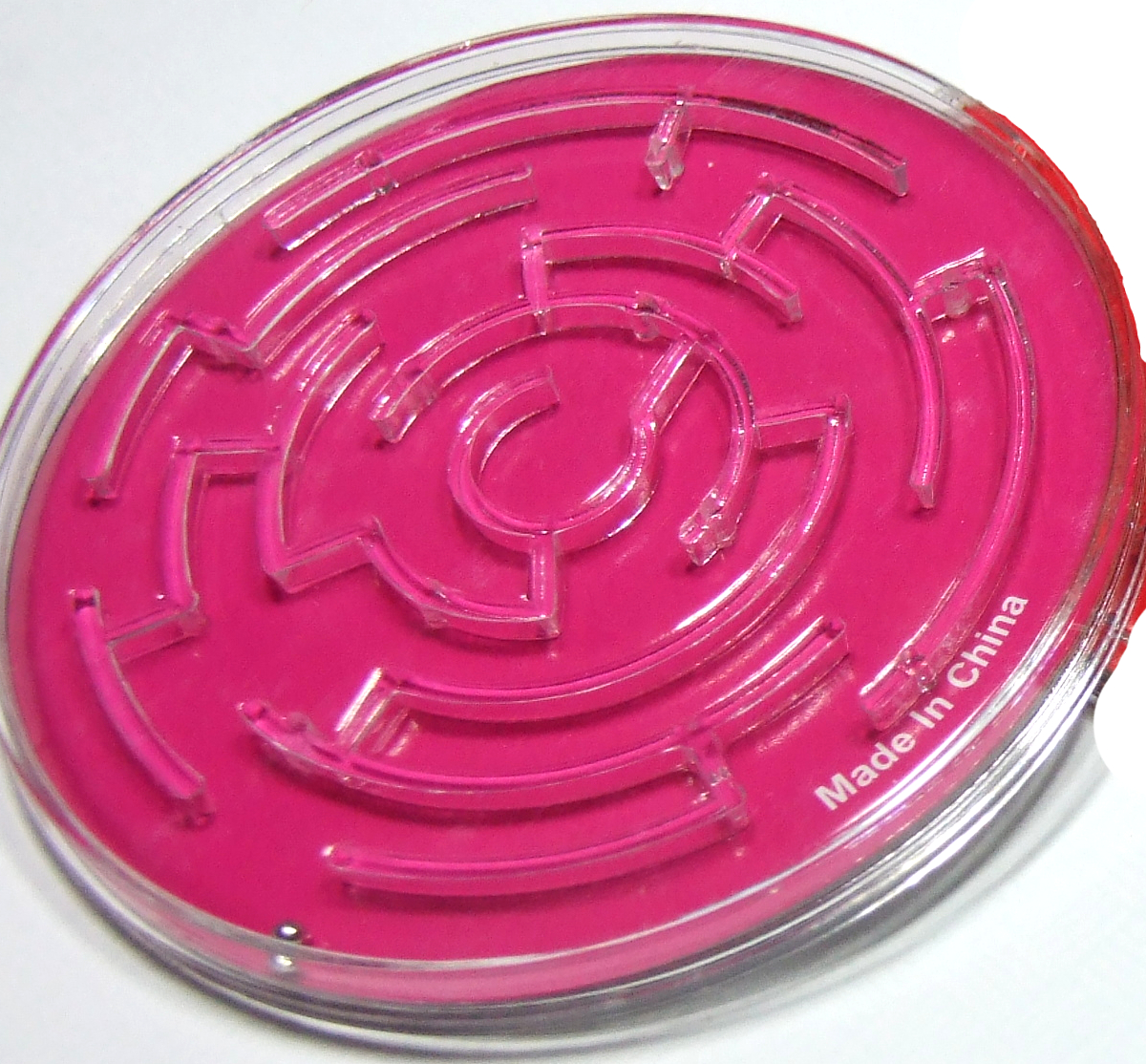}}
\subfloat[]{\includegraphics[width=0.49\textwidth]{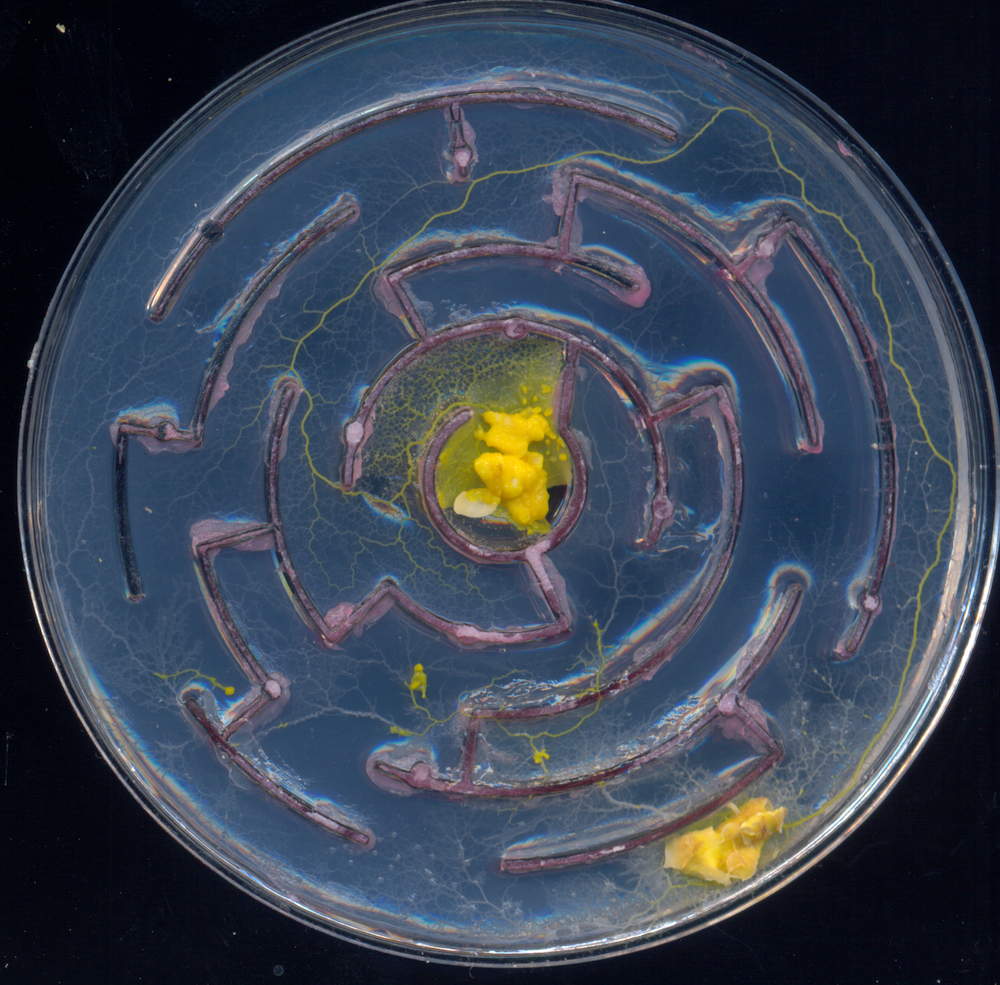}}\\
\subfloat[]{\includegraphics[width=0.49\textwidth]{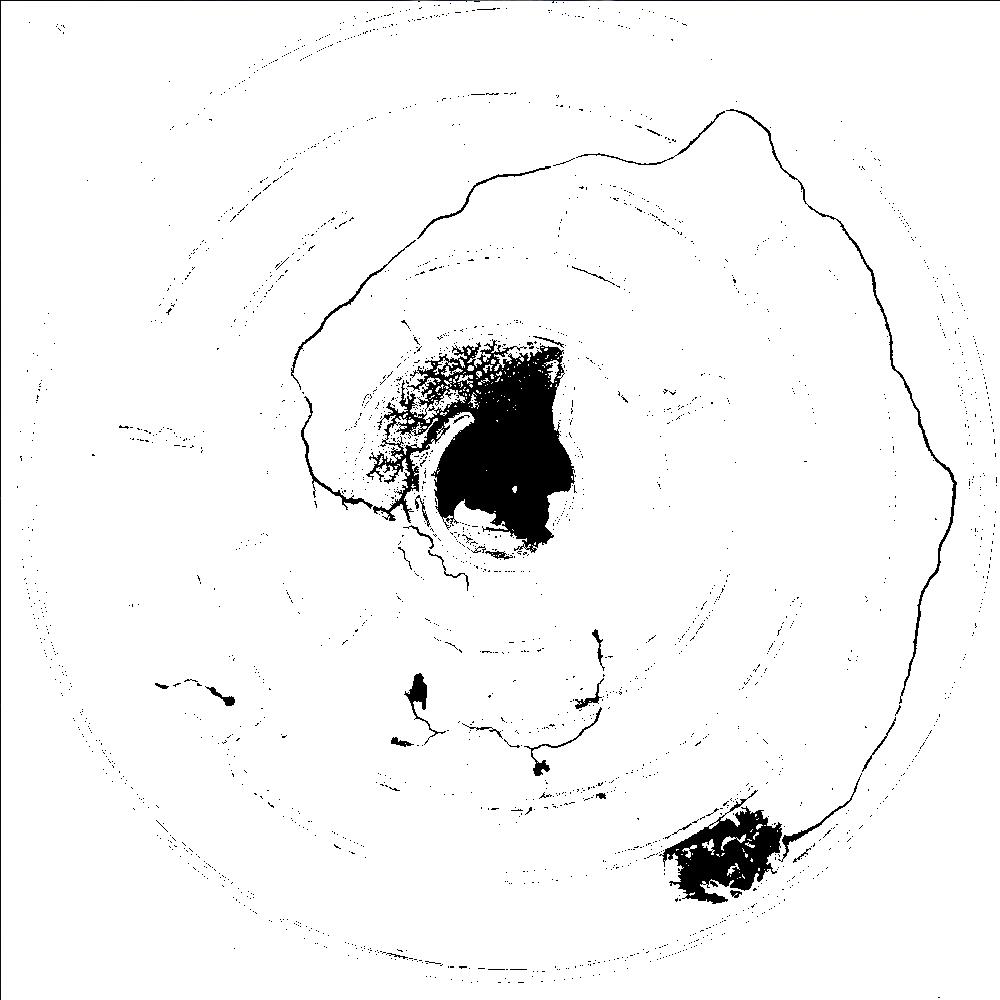}}
\subfloat[]{\includegraphics[width=0.49\textwidth]{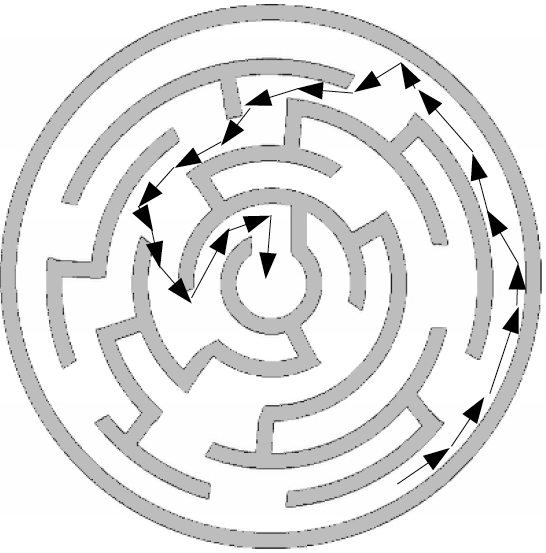}}
\caption{Experimental maze-solving with plasmodium of \Physarum:
(a)~maze used in experiments on slime mould maze-solving;
(bc)~plasmodium is inoculated in peripheral channel, east part of the maze, and a virgin oat 
flake is place in central chamber;
(b)~scanned image of the experimental maze, protoplasmic tubes are yellow (light gray); 
(c)~binarised, based on red and green components, images, major protoplasmic tubes are thick black lines;
(d)~scheme of plasmodium propagation, arrows symbolise velocity vectors of propagating active zone.
See experimental laboratory videos at \url{http://www.youtube.com/user/PhysarumMachines}}
\label{experiment01}
\end{figure*}

In laboratory experiments we used plastic mazes (Tesco's Toy Mazes, Tesco Plc), 70~mm diameter with 4~mm wide  and 3~mm deep channels (Fig.~\ref{experiment01}a).  We filled channels with 2\% agar gel (Select agar, Sigma Aldrich) as a non-nutrient substrate.  We smeared top of channel walls with strawberry flavoured Chasptick (Pfizer Consumer Healthcare Ltd) to deter plasmodium from making 'illegal' shortcuts over the walls separating channels.
A rolled oat was placed in the central chamber of the maze and an oat flake colonised by plasmodium of \Physarum was placed in the most peripheral channel of the maze. Mazes with plasmodium were kept in the dark in 20-23~C$^o$ temperature. Images of mazes were scanned in Epson Perfection 4490. Photos are taken using FujiPix 6000 camera.

A typical experiment is illustrated in Fig.~\ref{experiment01}. We placed an oat flake in the 
central chamber and inoculated plasmodium of \Physarum in a peripheral channels. The plasmodium started 
exploring its vicinity and at first generated two active zones propagating clock- and contra-clockwise. By the time diffusing chemo-attractants reached distant channels one of the active zone already became dominant and suppressed another active zone. In example  shown in Fig.~\ref{experiment01}ab active zone travelling contra-clockwise dominated
and 'extinguished' active zone propagating clockwise. The dominating active zone then  followed gradient of chemo-attractants inside the maze, navigated along intersections of the maze's channels and solved the maze by entering its central chamber.

\section{Mechanics of maze-solving: Numerical modelling}

A plasmodium of \Physarum can be seen as a network of coupled bio-chemical 
oscillators~\cite{takahashi_1997,tero_2005}. Interactions between the oscillators determine space-time dynamics of
contractile activity and protoplasmic streaming in protoplasmic 
tubes~\cite{kamiya_1959,achenbach_1980} and ultimately shape of plasmodium's cell~\cite{nakagaki_2000}.
Laboratory experiments show that oscillators in plasmodium network interact similarly to neurons in a simple neural networks~\cite{takamatsu_2006}. We can speculate that plasmodium's active zones (analogs of growth cones of 
maturing neuroblast) establish mutually inhibiting relationships. An active zone proximal to a source of chemo-attractants 
sends biochemical and electrical signals which suppress activity of active zones distal to the source.  

\begin{figure*}[!t]
\centering
\subfloat[]{\includegraphics[width=0.49\textwidth]{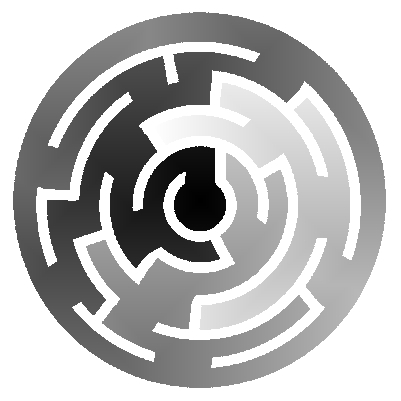}}
\subfloat[]{\includegraphics[width=0.49\textwidth]{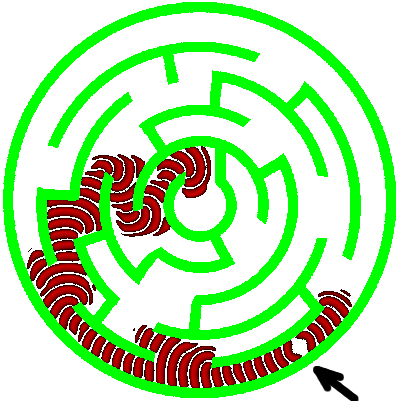}}\\
\subfloat[]{\includegraphics[width=0.49\textwidth]{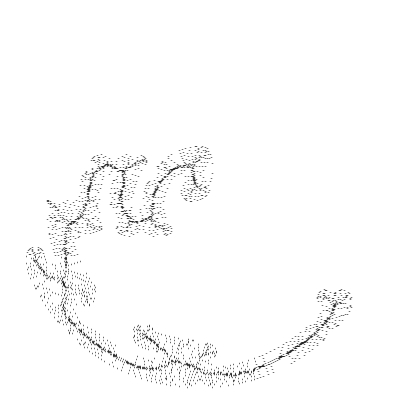}}
\caption{Simulating maze-solving in Oregonator model: (a)~gradient of chemo-attractants originated in central chamber, intensity of black is proportional to concentration of the attractants; impassable walls of the maze are white; 
(b)~time-lapse images of propagating active zones, active zones are recorded every 400th step of simulation as sites with $u_x>0.1$ are shown by red (dark gray), impassable walls of the maze are green (light gray); site of initial medium's perturbation, analog of plasmodium inoculation, is indicated by arrow;
 (c)~major network of protoplasmic tubes extracted from (b).  See videos at \url{http://www.youtube.com/user/PhysarumMachines}}
\label{maze400}
\end{figure*}

A profile of plasmodium's active zone on a non-nutrient substrate is isomorphic to shapes of and behaves analogously to  wave-fragments in sub-excitable media~\cite{adamatzky_physarummachines}. When active zone propagates two processes occur simultaneously --- movement of the wave-shaped tip of the pseudopodium and formation of a trail of protoplasmic tubes.  We simulate the tactic traveling of plasmodium growth front using two-variable Oregonator equation~\cite{field_noyes_1974}:
$$\frac{\partial u_x}{\partial t} = \frac{1}{\epsilon} (u_x - u_x^2 - (f v_x + \phi_x)\frac{u_x-q}{u_x+q}) + D_u \nabla^2 u$$
$$\frac{\partial v_x}{\partial t} = u_x - v_x .$$
The variable $u_x$ is abstracted as a local density of plasmodium's protoplasm at site $x$ 
and $v_x$ reflects local concentration of 
metabolites and nutrients. Parameters $q$ and $f$ are inherited from model of Belousov-Zhabotinsky medium. Parameter $\phi_x$ characterises excitability of medium in the Oregonator model and can be seen as analog of a degree of plasmodium's isometric tension response to chemo-attractants' concentration~\cite{ueda_1976}.
 We integrate the system using Euler method with five-node Laplacian, time step $\Delta t=5\cdot10^{-3}$ and grid point spacing $\Delta x = 0.25$,  $\epsilon=0.03$, $f=1.4$, $q=0.022$. Walls of the maze are impassable obstacles, grid nodes occupied by the walls are non-excitable.

Let $c_x$ be a concentration of chemo-attractant at site $x$. At moment $t=0$
only site occupied by oat flake has concentration $c_x=1$, all other sites have zero concentration. At every moment
$t$ of simulation a site $x$ updates its state $c_x$ by the following rule. If $c_x^t=0$ and there is at least one immediate neighbour $y$ such that $c_y>0$ then $c_x^{t+1}=t^{-1}$. Such rude approximation of diffusion is enough to build satisfactory gradient of chemo-attractants for guiding propagating wave-fronts. The gradient developed in simulation is shown in Fig.~\ref{maze400}a.

Initially all sites have the same excitability parameter $\phi^0_x=0.04515$. The less is value of $\phi_x$  
the higher is excitability of $x$. The excitability is revised as follows. At every $\sigma$th step of simulation we detect excited, i.e. $u_z \geq 0.1$, site $z$ with maximum, amongst all sites excited at this moment, concentration of chemo-attractants $c_z$. Then for every site $x$ we revise its excitability as follows: if $c_x < c_z$ then $\phi_x=0.09$.  In simulation illustrated in  Fig.~\ref{maze400}b $\sigma=500$. The medium is perturbed by an initial excitation, where a $5\times5$ sites are assigned $u=1.0$ each (Fig.~\ref{maze400}b). The perturbation generates a propagating wave-fragment travelling along gradient of chemo-attractants. Some travelling waves branch but branches are extinguished due to increase of sensitivity threshold.

 To imitate formation of protoplasmic tubes we store values of $u$ in matrix $\bf L$, which is processed at the end of simulation. For any site $x$ and time step $t$ if $u_x>0.1$ and $L_x=0$ then $L_x=1$. The matrix $\bf L$ represents time lapse superposition of propagating wave-fronts. The simulation is considered completed when propagating pattern reaches destination site (central chamber in scenario illustrated in Fig.~\ref{maze400}) and halts any further motion.  At the end of simulation we repeatedly apply the erosion operation~\cite{adamatzky_physarummachines}  to $\bf L$. This operation symbolises  a stretch-activation effect~\cite{kamiya_1959} necessary for formation of plasmodium tubes. The resultant protoplasmic network  provides a good phenomenological match for networks recorded in laboratory experiments (Fig.~\ref{maze400}c).

\section{Discussion}

\begin{figure*}[!t]
\centering
\subfloat[]{\includegraphics[width=0.49\textwidth]{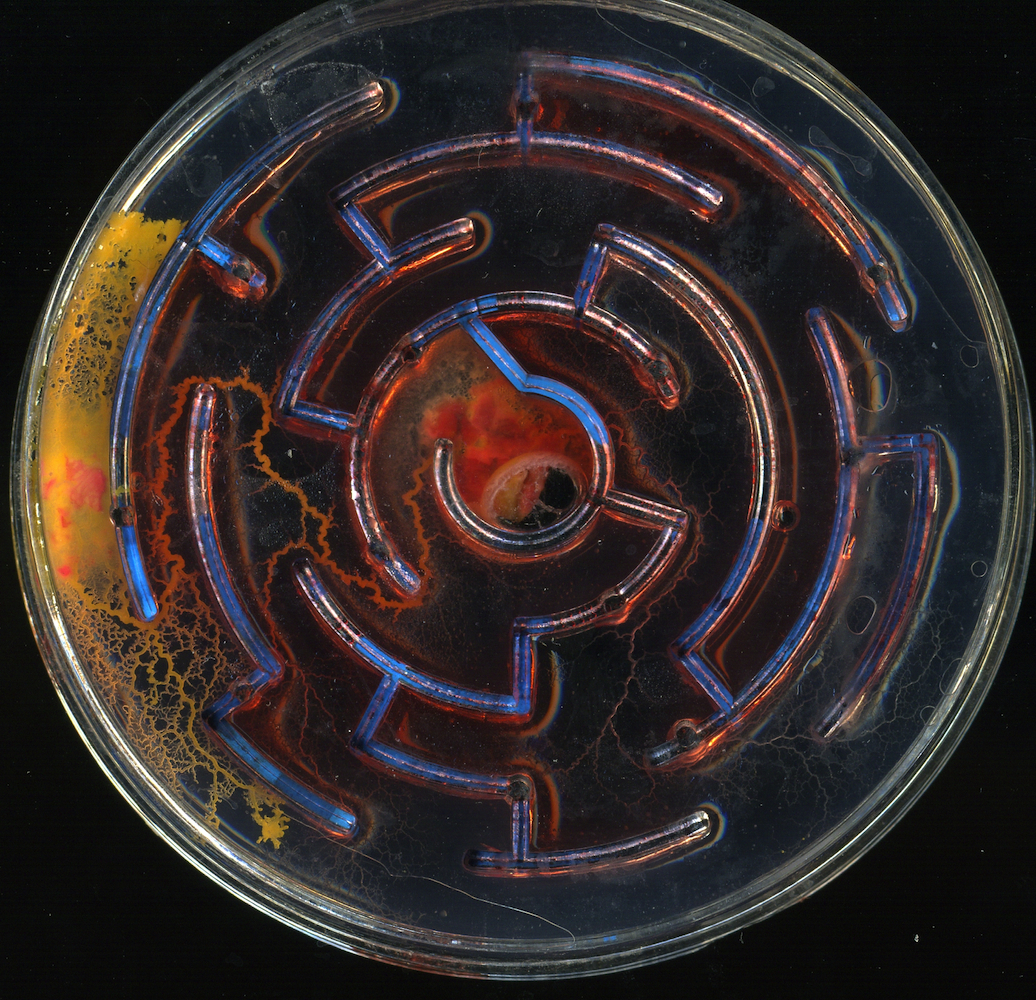}}
\subfloat[]{\includegraphics[width=0.49\textwidth]{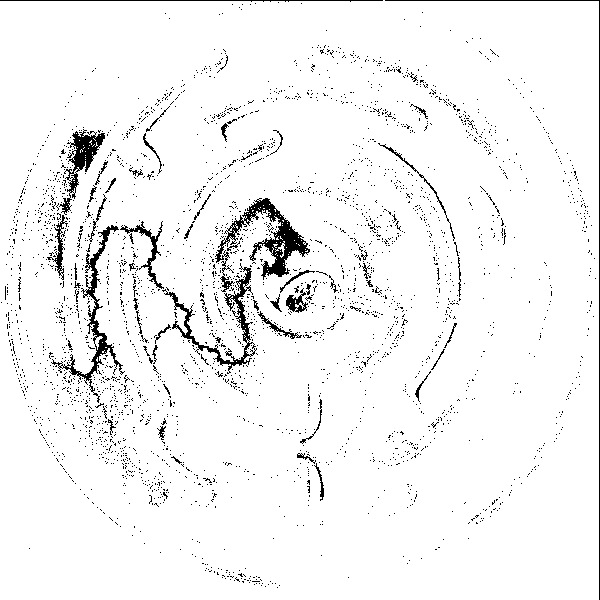}}
\caption{Experimental maze-solving with plasmodium of \Physarum:~plasmodium is inoculated in central chamber and a virgin flake is place in the most peripheral channel, west of the maze. (a)~photo of experimental maze, (b)~binarised image. }
\label{centralchamber}
\end{figure*}

We experimentally demonstrated that plasmodium of \emph{P. polycephalum} solves maze in one-pass, i.e. without exploring all possible solutions, if a source of chemo-attractants is placed at destination site.  
In our experiments we inoculated slime mould in a peripheral channel and a target oat flake in central chamber.  Positions do not matter, if there is an obstacle-free pass from a source to a destination the slime mould will trace it. For example, 
slime mould inoculated in a central chamber finds exit out of a maze. In Fig.~\ref{centralchamber} we placed an oat flake colonised by plasmodium in the central chamber and put an attracting oat flake in the west part of an outer channel. The plasmodium navigated thought first junction. It branched at the second junction. Active zone travelling clock-wise detected higher concentration of chemo-attractants than concentration detected by contra-clockwise propagating active zone. Thus active zone moving clock-wise became dominating. Eventually it reached its destination site (oat flake) along the shortest path (Fig.~\ref{centralchamber}).

\begin{table}[!t]
\caption{Brief comparison of laboratory prototypes of maze-solvers implemented in spatially extended physical, chemical or biological media. Two stages of maze-solving are outlined: maze exploration, or computation of many-sources-one-destination paths, and path tracing, or following from  source site to specified destination site.}
\begin{tabular}{lll}
Prototype 			&	Maze is explored by  		&	Path is traced by  \\ \hline
				&						&				\\
BZ medium~\cite{steinbock_1995}			& 	Excitation waves			& 	Computer		\\ 
Pruning plasmodium~\cite{nakagaki_2000}	&	Plasmodium				&	Plasmodium		\\
Gas-discharge~\cite{reyes_2002}	& Electrical field 				&  Electrical field	\\
Hot ice~\cite{adamatzky_hotice} 			&	Crystallisation pattern		&       Computer		\\
Mobile droplet~\cite{lagzi_2010}		&	Diffusing chemicals 		&	Droplet       		\\
Single-pass plasmodium	&	Diffusing chemo-attractants	&	Plasmodium		\\
\end{tabular}
\label{comparisontable}
\end{table}

\begin{figure*}[!tbp]
\centering
\subfloat[]{\includegraphics[width=0.49\textwidth]{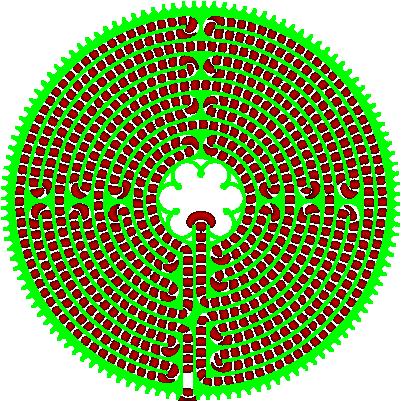}}
\subfloat[]{\includegraphics[width=0.49\textwidth]{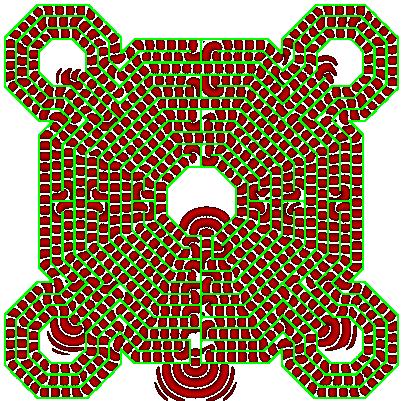}}
\caption{Solving Chartres~(a) and Reims~(b) mazes in Oregonator model of slime mould. 
Time-lapse images of an active zones (analogs of excitation wave-fronts in sub-excitable chemical medium) 
following a gradient of chemo-attractants emitted by central chamber of mazes. At the beginning of experiments
simulated plasmodium is inoculated at the mazes' entrances, south edges of mazes. }
\label{chartresreims}
\end{figure*}

Is our prototype better --- 
in terms of complexity or real-life speed and costs --- then other laboratory prototypes of maze-solvers (Tab.~\ref{comparisontable})? All existing prototypes have the same computational time complexity $O(L)$, where $L$ is a length of a worst-case scenario shortest path in a maze (a worst case is when shortest path spans all channels of a maze, as e.g. Chartres and Remis mazes shown in Fig.~\ref{chartresreims}).

The time complexity is determined by time taken by waves, electricity, plasmodium or diffusing chemicals to explore the maze. Al these substances explore maze in parallel by propagating from their original locations  to all other sites of maze, thus it takes them $O(L)$ steps to span the maze. Tracing of a path from source to destination is rather a menial tasks --- to follow gradients or other physical or chemical characteristics of the medium. It takes $O(L)$ steps as well. Space complexity of all prototypes is the same $O(L)$: it takes $O(L)$ instances (molecules, charges, micro-volumes of 
mixture or protoplasm) to fill in all channels during exploration stage. 

In terms of a real time gas-discharge~\cite{reyes_2002} is the fastest maze-solver, it takes just hundred of milliseconds to solve the maze. Hot ice computer~\cite{adamatzky_hotice} is the second fastest one, a maze of 0.1~m in diameter can be solved in few seconds. Gas-discharge and hot-ice solvers are followed by BZ-medium~\cite{steinbock_1995}	 and mobile-droplet~\cite{lagzi_2010} solvers: solution time is measured in minutes and hours. Slime mould based solvers are the slowest ones: it usually takes them a couple of days to solve a maze like one we used in present experiments.  The prototypes can be arranged in the following real-costs (which includes consumables and laboratory equipment) descending order: gas-discharge, mobile droplets, BZ-medium, hot ice, and slime mould. Gas-discharge maze solver is most expensive while \Physarum solver can run literally for free. 

\begin{figure*}[!t]
\centering
\subfloat[]{\includegraphics[width=0.49\textwidth]{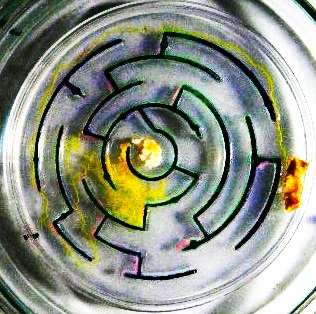}}
\subfloat[]{\includegraphics[width=0.49\textwidth]{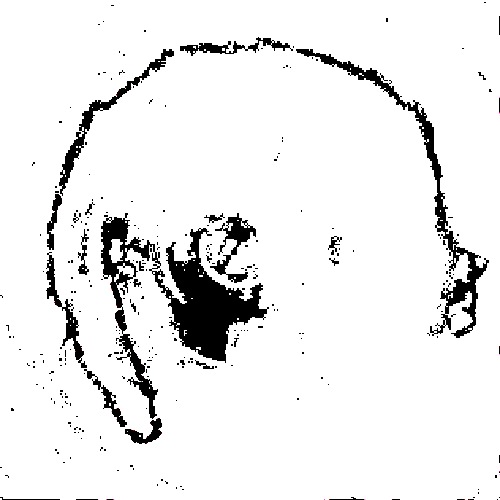}}
\caption{Plasmodium of \Physarum chooses longest path to maze's central chamber: 
plasmodium is inoculated in peripheral channel, east part of the maze, and a virgin oat 
flake is place in central chamber;
(a)~scanned image of the experimental maze, protoplasmic tubes are yellow (light gray); 
(b)~binarised, based on red and green components, images, major protoplasmic tubes are thick black lines.}
\label{experiment02}
\end{figure*}

Accuracy of slime mould computing is far from ideal.  Situation illustrated in Fig.~\ref{experiment01} is a typical one: slime computes almost shortest path, compare with the shortest path from source to destination computed in Oregonator model in  Fig.~\ref{maze400}. In some situations, e.g. the one shown in  Fig.~\ref{experiment02}, plasmodium chooses the longest path from source to destination. Said that in neither of 35 experiments we undertook plasmodium failed to solve the maze, it solves builds a path from source to destination. 

Our final 'disclaimer' is that neither of slime mould  maze-solvers can successfully compete with existing conventional computer architectures. However, \emph{P. polycephalum} is an ideal biological substrate which represents all essential features of reaction-diffusion chemical computers yet encapsulated in an elastic growing membrane~\cite{adamatzky_physarummachines}. 
It can be treated as a meso-scale prototype of future growing nano-scale circuits. We envisage slime mould maze-solvers can also contribute toward designs of parallel amorphous drug delivery systems, or smart needles, which grow towards the target tissue.

\end{document}